\documentclass[onecolumn,showpacs,preprintnumbers,showkeys]{revtex4}
\usepackage{amsthm, amscd, amsfonts, amssymb, graphicx}
\usepackage{dcolumn}
\usepackage{bm}
\usepackage[english]{babel}

\begin{document}

\title{Effective Ginzburg-Landau free energy functional for multi-band isotropic superconductors}
\author{Konstantin V. Grigorishin}
\email{konst.phys@gmail.com}
 \affiliation{Bogolyubov Institute for
Theoretical Physics of the National Academy of Sciences of
Ukraine, 14-b Metrolohichna str. Kiev-03680, Ukraine.}
\date{\today}

\begin{abstract}

It has been shown that interband mixing of gradients of two order
parameters (drag effect) in an isotropic bulk two-band
superconductor plays important role - such a quantity of the
intergradients coupling exists that the two-band superconductor is
characterized with a single coherence length and a single
Ginzburg-Landau (GL) parameter. Other quantities or neglecting of
the drag effect lead to existence of two coherence lengths and
dynamical instability due to violation of the phase relations
between the order parameters. Thus so-called \textit{type-1.5}
superconductors are impossible. An approximate method for solving
of set of GL equations for a multi-band superconductor has been
developed: using the result about the drag effect it has been
shown that the free-energy functional for a multi-band
superconductor can be reduced to the GL functional for an
effective single-band superconductor.
\end{abstract}


 \maketitle

\section{Introduction}\label{intr}

Two-band superconductors are a specific class of superconductors
essentially differing in their properties from single-band
superconductors. Their typical representatives are magnesium
diboride $\texttt{MgB}_{2}$, strontium ruthenate
$\texttt{Sr}_{2}\texttt{Ru}\texttt{O}_{4}$, nonmagnetic
borocarbides $\texttt{LuNi}_{2}\texttt{B}_{2}\texttt{C}$,
$\texttt{YNi}_{2}\texttt{B}_{2}\texttt{C}$ and ferropnictides. In
this article we will consider only isotropic bulk
(polycrystalline) s-wave superconductors. One of the main feature
of these materials is the presence of two energy gaps $\Delta_{1}$
and $\Delta_{2}$ which, however, vanishes at the same temperature
$T_{c}$ (Fig.\ref{Fig1}). According to microscopic theory
\cite{krist,ciech,dol,litak} presence of the two gaps is explained
by the fact that in each band $i$ an own coupling constant
$g_{ii}$ exists - the intraband constant. In the same time, the
interband coupling constant $g_{ij}$ exists too, which, on the one
hand, enhances pairing of electrons, on the other hand, leads to
the single critical temperature $T_{c}$. BCS gap equations for a two-band superconductor are 
\cite{krist,litak,dol}:
\begin{eqnarray}\label{0.1}
&&\Delta_{1}=\sum_{\textbf{k}}\frac{g_{11}\Delta_{1}\tanh(E_{1,k}/2k_{\texttt{B}}T)}{2E_{1,k}}
+\sum_{\textbf{k}}\frac{g_{12}\Delta_{2}\tanh(E_{2,k}/2k_{\texttt{B}}T)}{2E_{2,k}}\nonumber\\
&&\Delta_{2}=\sum_{\textbf{k}}\frac{g_{22}\Delta_{2}\tanh(E_{2,k}/2k_{\texttt{B}}T)}{2E_{2,k}}
+\sum_{\textbf{k}}\frac{g_{12}\Delta_{1}\tanh(E_{1,k}/2k_{\texttt{B}}T)}{2E_{1,k}},
\end{eqnarray}
where $E_{i,k}$ is the quasiparticle's energy in a band $i$.
Unlike single-band BCS theory a superconducting state can exist
both attractive interband coupling constant $g_{12}>0$ and
repulsive $g_{12}<0$, moreover the gaps are nonzero if even the
intraband couplings are absent $g_{11}=g_{22}=0$. In the case of
the attractive interband interaction the gaps have the same phases
on both Fermi surfaces, while for the repulsive interaction the
phases will be opposite. Thus the phase difference of the order
parameters $|\Delta_{1}|e^{i\varphi_{1}},
|\Delta_{2}|e^{i\varphi_{2}}$ are:
\begin{equation}\label{0.2}
    \begin{array}{cc}
      \cos(\varphi_{1}-\varphi_{2})=1 & \texttt{if}\quad g_{12}>0  \\
      \cos(\varphi_{1}-\varphi_{2})=-1 & \texttt{if}\quad g_{12}<0 \\
    \end{array}
\end{equation}
For example, in absence of magnetic field we can suppose
$\Delta_{1}>0$, then we will have $\Delta_{2}>0$ for $g_{12}>0$
and $\Delta_{2}<0$ for $g_{12}<0$. From Eq.(\ref{0.1}) we can see
the important property of a two-band superconductor: if we violate
the phase relation (\ref{0.2}) then suppression of the energy gaps
$\Delta_{1},\Delta_{2}$ will take place (extremely strong
suppression if the intraband couplings are absent
$g_{11}=g_{22}=0$). For the suppression of the order parameters
the violation of the phase-locked states
$\varphi\equiv\varphi_{1}-\varphi_{2}=0$ or $\pi$ must be
macroscopic, constant in time and not small (for example, when
there are two different coherence lengths in a system with two
gaps), unlike Leggett's mode, which is collective mode of small
fluctuations of relative phase $\varphi(\textbf{r},t)$ and behaves
like the Anderson plasmons in Josephson junctions
\cite{legg,shar}. In addition, we assume that current in a
two-band superconductor is less than some a critical current
$J<J_{t}$ over which interband phase breakdown occurs, resulting
in spontaneous phase solitons in $\varphi(\textbf{r},t)$
\cite{gur1}, which is nonequilibrium state. In phenomenological
theory the coupling between the bands is represented by
Josephson-like coupling term:
\begin{equation}\label{0.3}
  \varepsilon\left(\Psi_{1}^{+}\Psi_{2}+\Psi_{1}\Psi_{2}^{+}\right)
\end{equation}
in a free energy functional, where $\Psi_{1}$ and $\Psi_{2}$ are
order parameters for band 1 and 2 accordingly.

Currently there are two opinions about the properties of two-band
superconductors:

1) In papers \cite{mosh,nish} it has manifested about a new type
of superconductivity in $\texttt{MgB}_{2}$ - a novel "type-1.5
superconductor", contrary to type-I and type-II superconductors.
In papers \cite{bab1,bab2,bab3} a two-band superconductor was
studied, where they considered GL parameters
$\kappa_{i}=\lambda/\xi_{i}$ ($i=1, 2$) in two different regimes
to produce type-I ($\kappa_{1}< 1/\sqrt{2}$) and type-II
($\kappa_{2}>1/\sqrt{2}$) materials, that corresponds to different
coherence lengths $\xi_{1}=\frac{\hbar v_{F1}}{\pi\Delta_{1}(0)}$
and $\xi_{2}=\frac{\hbar v_{F2}}{\pi\Delta_{2}(0)}$. That is each
correlation length is sorted with a corresponding band, where
Fermi velocities $v_{F1},v_{F2}$ and energy gaps
$\Delta_{1},\Delta_{2}$ are different. Their prediction leads to
what they call a "semi-Meissner state". Instead of homogeneous
distribution, the vortexes form aperiodic clusters or vortexless
Meissner domains, arising out of short range repulsion and long
range attraction between vortexes.

2) However, in review \cite{brant} an opposite opinion has been
suggested in respect of existence of the type-1.5
superconductivity in two-band superconductors. It was shown that
for the real superconductor $\texttt{MgB}_{2}$ which possesses a
single transition temperature, the assumption of two independent
order parameters with separate penetration depths and separate
coherence lengths is unphysical. In particular, in the
above-mentioned works \cite{mosh,nish} numerical estimates for
$\xi_{i}$ are obtained by using the one-band BCS formula. On the
other hand, in works \cite{litak,ord} it has been shown that in a
two-band superconductor there are two coherence lengths which are
not related to the concrete bands involved in the formation of the
superconducting ordering in a system with interband interaction:
one of the lengths is diverges at the critical temperature
$\xi_{1}(T\rightarrow T_{c})\rightarrow\infty$, the second of them
is a nearly constant at all temperatures $\xi_{2}(T)\approx
const$. Besides it is necessary to be more accuracy at
calculations of interaction between vortexes - many corrections to
the simple GL or London theories are expected to modify the
monotonically decreasing interaction potential at large distances,
$V(r)\propto exp(-r/a)$, such that $a$ becomes complex. This, in
principle, causes an oscillating potential, whose first minimum
may occur at large distances where the amplitude of the potential
is small. Generally, as discussed in \cite{brant}, it should be
taken into account dependence on the material, its purity,
magnetic history, and temperature. In a paper \cite{kog} it was
shown that coherence length is the same for both order parameters
$\Delta_{1}, \Delta_{2}$, moreover the ratio of the order
parameters is $T$-independent in the GL domain,
$\Delta_{1}(\textbf{r},T)/\Delta_{2}(\textbf{r},T)=const$, with
the constant depending on interactions responsible for
superconductivity - thus the type-1.5 superconductivity is absent.
In a paper \cite{geyer} it was demonstrated that close to the
transition temperature, where the GL theory is applicable, the
two-band problem maps onto an effective single-band problem with a
GL parameter $\kappa^{-2}=\kappa^{-2}_{1}+\kappa^{-2}_{2}$, a
penetration depth $\lambda^{-2}=\lambda^{-2}_{1}+\lambda^{-2}_{2}$
and a coherence length $\xi=(\xi^{-2}_{1}+\xi^{-2}_{2})^{-1/2}$
where $\kappa_{i},\lambda_{i},\xi_{i}$ are quantities
corresponding to a band $i$. Similar effective single-band GL
approach also was applied in papers \cite{asker0,asker4}. The
two-band GL theory has been developed in works
\cite{asker1,asker2,asker3} where it was shown that the presence
of two order parameters leads to a nonlinear temperature
dependence of the upper and lower critical fields
$H_{c2}(T),H_{c1}(T)$ and thermodynamic magnetic field $H_{cm}(T)$
unlike single-band GL theory. In \cite{asker6} the temperature
dependence of the London penetration depth $\lambda(T)$ has been
determined. These results are in good agreement with the
experimental data for bulk $\texttt{MgB}_{2}$ and borocarbides
without any hypothesis about "type-1.5 superconductor" and
"semi-Meissner state".

In this paper we study two problems which, in our opinion, are
important for GL theory of isotropic bulk multi-band
superconductors:

1) The coupling between the bands is represented by both the term
of proximity effect Eq.(\ref{0.3}) and the term of drag effect -
interband mixing of order parameters' gradients:
\begin{equation}\label{0.4}
    \eta\left(\nabla\Psi_{1}^{+}\nabla\Psi_{2}+\nabla\Psi_{1}\nabla\Psi_{2}^{+}\right).
\end{equation}
Since electron from different bands are interacting, hence, if in
some a band the order parameter is spatially inhomogeneous
$\Psi_{1}(\textbf{r})$ then in other band the order parameter must
be inhomogeneous too $\Psi_{2}=\Psi_{2}(\textbf{r})$. If a current
exists in one band then it drags Cooper pairs in other band.
Therefore the coefficient $\eta$ must be function of carriers'
mass in each band $m_{1}$, $m_{2}$ and the coupling $\varepsilon$
between the order parameters. As a rule the drag effect is
neglected or the coefficient $\eta$ is considered as an adjustable
parameter. However in a work \cite{yerin}, where they considered
Little-Parks effect for two-band superconductors, it has been
found that the coefficient $\eta$ is not a arbitrary quantity and
a relation between the coefficient and effective masses of
carriers exists to ensure the existence of the absolute minimum of
the free energy functional. In present paper we show that the drag
effect plays important role in two-band superconductors.
Accounting of the drag effect leads to single coherence length
$\xi$ for a two-band superconductor unlike the papers
\cite{litak,ord}. Moreover the ratio of the order parameters is
$T$-dependent
$\Delta_{1}(\textbf{r},T)/\Delta_{2}(\textbf{r},T)=const(T)$,
unlike the work \cite{kog}. Neglecting of the drag effect leads to
dynamical instability of the two-band superconductor due to
violation of the phase relations (\ref{0.2}). Thus type-1.5
superconductors are impossible. Unlike previous works we have
found the coefficient $\eta$ as a function of $m_{1}$,
$m_{2},\varepsilon$.

2) GL equation for a single-band superconductor (in absence of a
magnetic field) is a nonlinear second-order differential equation.
Phenomenological theory for bulk isotropic two-band
superconductors has been developed in works
\cite{asker1,asker2,asker3}, where GL equations are a set of two
nonlinear second-order differential equations. Exact GL theory for
two-band superconductors is mathematical complicated and
cumbersome. Generalized set of GL equations for multi-band
superconductors will be extremely complicated. Therefore
approximate methods are required. In this paper we show that,
using the result about the drag effect, the GL theory for a
two-band superconductor can be reduced to the GL theory for an
effective single-band superconductor. Generalizing this result we
develop an algorithm which allows to reduce the free energy
functional of a multi-band superconductor to the GL free energy
functional of an effective single-band superconductor.

\section{Two-band superconductor}\label{two}

In presence of two-order parameters in a bulk isotropic s-wave
superconductor, the GL free energy functional can be written as
\cite{asker6,asker1,asker2,asker3,doh}:
\begin{eqnarray}\label{1.1}
    &&F=\int d^{3}r[\frac{\hbar^{2}}{4m_{1}}\left|D\Psi_{1}\right|^{2}
    +\frac{\hbar^{2}}{4m_{2}}\left|D\Psi_{2}\right|^{2}+
    \frac{\hbar^{2}}{4}\eta\left(D^{+}\Psi_{1}^{+}D\Psi_{2}
    +D\Psi_{1}D^{+}\Psi_{2}^{+}\right)\nonumber\\
    &&+a_{1}\left|\Psi_{1}\right|^{2}+a_{2}\left|\Psi_{2}\right|^{2}
    +\frac{b_{1}}{2}\left|\Psi_{1}\right|^{4}+\frac{b_{2}}{2}\left|\Psi_{2}\right|^{4}
    +\varepsilon\left(\Psi_{1}^{+}\Psi_{2}+\Psi_{1}\Psi_{2}^{+}\right)+\frac{H^{2}}{8\pi}],
\end{eqnarray}
where the differential operator are $D=\nabla-\frac{2\pi
i}{\Phi_{0}}\textbf{A}$ ($\Phi_{0}=\pi\hbar c/e$ is a magnetic
flux quantum, $\textbf{H}=\texttt{rot}\textbf{A}$ is a vector
potential), $m_{1,2}$ denotes the effective mass of carriers in
the correspond band, the coefficient $a$ is given as
$a_{i}=\gamma_{i}(T-T_{ci})$, $\gamma$ is constant, the
coefficients $b_{1,2}$ are independent on temperature, the
quantities $\varepsilon$ and $\eta$ describe interband mixing of
two order parameters (proximity effect) and their gradients (drag
effect), respectively. If we switch off the interband interaction
$\varepsilon=\eta=0$ then we will have two independent
superconductors with the different critical temperatures $T_{c1}$
and $T_{c2}$ because the intraband interactions can be different
$g_{11}\neq g_{22}$. There is another form of the coefficients
$a_{i}$ \cite{zhit,kog}: they acquire constant parts
$const_{i}+\gamma_{i}(T-T_{c})$ such that
$const_{1}const_{2}=\varepsilon^{2}$ and $T_{c}$ is critical
temperature of a two-band superconductor. However in this case if
we switch off the interband interaction $\varepsilon=0\Rightarrow
const_{1,2}=0$, then we will have two independent superconductors
with the same critical temperatures $T_{c}$.

\begin{figure}[ht]
\includegraphics[width=7.0cm]{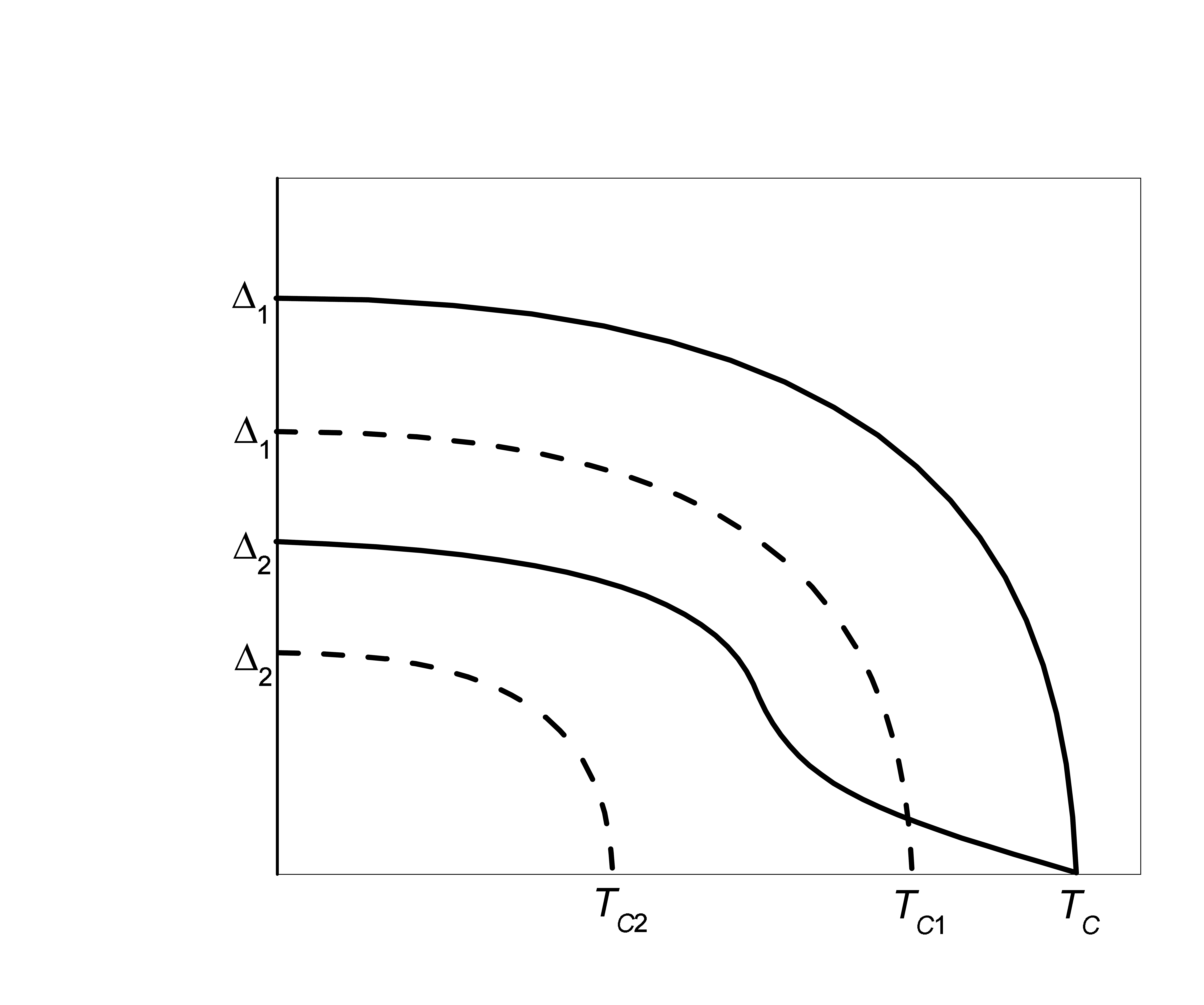}
\caption{Superconductor gap parameters $\Delta_{1}$ and
$\Delta_{2}$ if the interband interaction is absent
($\varepsilon=0$) (dash lines) and if the interband interaction
takes place ($\varepsilon\neq 0$) (solid line). } \label{Fig1}
\end{figure}

Minimization of the free energy functional with respect to the
order parameters, if $\nabla \Psi_{1,2}=0$ and $\textbf{A}=0$,
gives
\begin{equation}\label{1.2}
\left\{\begin{array}{c}
  a_{1}\Psi_{1}+\varepsilon\Psi_{2}+b_{1}\Psi_{1}^{3}=0 \\
  a_{2}\Psi_{2}+\varepsilon\Psi_{1}+b_{2}\Psi_{2}^{3}=0 \\
\end{array}\right\}
\end{equation}
Near critical temperature $T_{c}$ we have
$\Psi_{1,2}^{3}\rightarrow 0$, hence we can find the critical
temperature as a solvability condition of the linearized
Eqs.(\ref{1.2}):
\begin{equation}\label{1.3}
a_{1}a_{2}-\varepsilon^{2}=\gamma_{1}\gamma_{2}(T_{c}-T_{c1})(T_{c}-T_{c2})-\varepsilon^{2}=0.
\end{equation}
Solving this equation we find $T_{c}>T_{c1},T_{c2}$, moreover the
solution does not depend on sign of $\varepsilon$. The sign
determines the phase difference of the order parameters
$|\Psi_{1}|e^{i\varphi_{1}}, |\Psi_{2}|e^{i\varphi_{2}}$:
\begin{equation}\label{1.3a}
    \begin{array}{cc}
      \cos(\varphi_{1}-\varphi_{2})=1 & \texttt{if}\quad\varepsilon<0  \\
      \cos(\varphi_{1}-\varphi_{2})=-1 & \texttt{if}\quad\varepsilon>0 \\
    \end{array},
\end{equation}
that follows from the Eqs.(\ref{1.2},\ref{1.3}) and is an analogue
of Eq.(\ref{0.2}): the case $\varepsilon<0$ corresponds to
attractive interband interaction $g_{12}>0$, the case
$\varepsilon>0$ corresponds to repulsive interband interaction
$g_{12}<0$. It should be noted that the interband mixing of
two-order parameters $\varepsilon$ ensures the single critical
temperature $T_{c}$ of a two-band superconductor whilst each band
has own critical temperature - $T_{c1}$ and $T_{c2}$ if the
interband interaction is absent. This fact is illustrated in
Fig.(\ref{Fig1}), where it is given the qualitative picture of
calculations in \cite{litak}.

Phase relations (\ref{1.3a}) imposes restrictions on the
coefficient $\eta$. For temperatures near $T_{c}$ and magnetic
fields smaller than $H_{c1}$, the influence of the field on
modulus of the order parameters can be neglected and we assume
$|\Psi_{1}|=const,|\Psi_{2}|=const$. Then the wave function can be
written as $\Psi_{j}=|\Psi_{j}|\exp(i\varphi_{j}(\textbf{r}))$,
where $\varphi_{j}(\textbf{r})$ are the phases of the order
parameters. The GL free energy functional (\ref{1.1}) can be
rewritten as
\begin{eqnarray}\label{1.3b}
    &&F=\int d^{3}r[\frac{\hbar^{2}}{8m_{1}}n_{1}\left(\nabla\varphi_{1}-\frac{2\pi\textbf{A}}{\Phi_{0}}\right)^{2}
    +\frac{\hbar^{2}}{8m_{2}}n_{2}\left(\nabla\varphi_{2}-\frac{2\pi\textbf{A}}{\Phi_{0}}\right)^{2}\nonumber\\
    &&+\frac{\hbar^{2}}{4}\eta\sqrt{n_{1}n_{2}}
    \left(\nabla\varphi_{1}-\frac{2\pi\textbf{A}}{\Phi_{0}}\right)\left(\nabla\varphi_{2}-\frac{2\pi\textbf{A}}{\Phi_{0}}\right)\cos(\varphi_{1}-\varphi_{2})\nonumber\\
    &&+\varepsilon\sqrt{n_{1}n_{2}}\cos(\varphi_{1}-\varphi_{2})+\frac{H^{2}}{8\pi}],
\end{eqnarray}
where $n_{1}=2|\Psi_{1}|^{2}$ and $n_{2}=2|\Psi_{2}|^{2}$ are the
densities of superconducting electrons for the corresponding
bands. Phase relations (\ref{1.3a}) must be satisfied over the
entire volume of a superconductor:
$\varphi_{1}(\textbf{r})-\varphi_{2}(\textbf{r})=const$, otherwise
superconducting state will be destroyed -
Eqs.(\ref{0.1},\ref{0.2}). Hence the phases must change equally:
\begin{equation}\label{1.3c}
\nabla\varphi_{1}(\textbf{r})=\nabla\varphi_{2}(\textbf{r}).
\end{equation}
Minimizing the free energy functional (\ref{1.3b}) with respect to
the vector potential $\textbf{A}$ we find the current
$\textbf{J}=\frac{c}{4\pi}\nabla\times\textbf{H}$:
\begin{eqnarray}\label{1.3c1}
    &&\textbf{J}=\frac{2\pi c}{\Phi_{0}}[\frac{\hbar^{2}}{4m_{1}}n_{1}\left(\nabla\varphi_{1}-\frac{2\pi\textbf{A}}{\Phi_{0}}\right)
    +\frac{\hbar^{2}}{4m_{2}}n_{2}\left(\nabla\varphi_{2}-\frac{2\pi\textbf{A}}{\Phi_{0}}\right)\nonumber\\
    &&+\frac{\hbar^{2}}{4}\eta\sqrt{n_{1}n_{2}}
    \left\{\left(\nabla\varphi_{1}-\frac{2\pi\textbf{A}}{\Phi_{0}}\right)+
    \left(\nabla\varphi_{2}-\frac{2\pi\textbf{A}}{\Phi_{0}}\right)\right\}\cos(\varphi_{1}-\varphi_{2})].
\end{eqnarray}
Let us consider a superconductor with an inner cavity. We
integrate Eq.(\ref{1.3c1}) along a closed path lying within the
superconductor around the cavity at a distance from the cavity's
surface larger than magnetic penetration depth $\lambda$. Hence on
the path we have $\textbf{J}=0$ and integral on the right-hand is
equal to zero. Then
\begin{eqnarray}\label{1.3c2}
    &&\left(\frac{n_{1}}{m_{1}}+\eta\sqrt{n_{1}n_{2}}
    \cos(\varphi_{1}-\varphi_{2})\right)\oint\nabla\varphi_{1}d\textbf{l}
    +\left(\frac{n_{2}}{m_{2}}+\eta\sqrt{n_{1}n_{2}}
    \cos(\varphi_{1}-\varphi_{2})\right)\oint\nabla\varphi_{2}d\textbf{l}\nonumber\\
&&=\frac{2\pi\Phi}{\Phi_{0}}\left(\frac{n_{1}}{m_{1}}+2\eta\sqrt{n_{1}n_{2}}
    \cos(\varphi_{1}-\varphi_{2})+\frac{n_{2}}{m_{2}}\right),
\end{eqnarray}
where $\Phi=\oint\textbf{A}d\textbf{l}$ is a magnetic flux. Taking
into account the functions $\varphi_{1}$ and $\varphi_{1}$ must be
single-valued
$\oint\nabla\varphi_{1}d\textbf{l}=\oint\nabla\varphi_{2}d\textbf{l}=2\pi
n$, we find that the magnetic flux through the cavity takes a
discrete series $\Phi=n\Phi_{0}$ like in single-band
superconductors \cite{asker1,yerin}.

Let us analyze the functional (\ref{1.3b}). The term
$\varepsilon\sqrt{n_{1}n_{2}}\cos(\varphi_{1}-\varphi_{2})<0$
always because Eq.(\ref{1.3a}). This lowers the free energy. For
stability of the superconducting state it is necessary that a
spatial inhomogeneity of the order parameters enlarges the free
energy. Since we have
$\nabla\varphi_{1}\nabla\varphi_{2}=(\nabla\varphi_{1})^{2}=(\nabla\varphi_{2})^{2}>0$
from Eq.(\ref{1.3c}) then the stability condition is
\begin{equation}\label{1.3d}
\frac{n_{1}}{m_{1}}+\frac{n_{2}}{m_{2}}+2\eta\sqrt{n_{1}n_{2}}\cos(\varphi_{1}-\varphi_{2})>0.
\end{equation}
From the Eq.(\ref{1.3c1}) we find the London penetration depth in
the following form
\begin{eqnarray}\label{1.3e}
    \lambda^{-2}(T)=\frac{4\pi e^{2}}{c^{2}}\left[\frac{n_{1}(T)}{m_{1}}
    +\frac{n_{2}(T)}{m_{2}}+2\eta\sqrt{n_{1}(T)n_{2}(T)}\cos(\varphi_{1}-\varphi_{2})\right].
\end{eqnarray}
From this formula we can see the condition (\ref{1.3d}) ensures
$\lambda^{2}(T)>0$ when $n_{1}(T),n_{2}(T)\neq 0$. Thus the
condition (\ref{1.3d}) restricts the possible quantities of the
parameter $\eta$. Let $\nabla\varphi_{1}=\nabla\varphi_{2}=0$,
that is a paramagnetic current is absent. Then the free energy
functional takes the form
\begin{equation}\label{1.3f}
    F=\frac{1}{8\pi}\int d^{3}r[\textbf{H}^{2}+\lambda^{2}\left(\texttt{rot}\textbf{H}\right)^{2}],
\end{equation}
Let the field $\textbf{H}_{0}$ is directed along the axis Oz and a
superconductor are in a halfspace $x>0$ then the magnetic field
within the superconductor are
$H(x)=H_{0}\exp\left(-x/\lambda\right)$. Substituting this field
in Eq.(\ref{1.3f}) and integrating we have the free energy per
unit of square:
\begin{equation}\label{1.3h}
    F=\frac{H_{0}^{2}}{8\pi}\lambda.
\end{equation}
We can see the smaller London penetration depth $\lambda$ the
smaller free energy. Then from Eq.(\ref{1.3e}) it follows that
such quantities of the parameter $\eta$, when
\begin{equation}\label{1.3i}
    \eta\cos(\varphi_{1}-\varphi_{2})>0\Rightarrow\eta\varepsilon<0,
\end{equation}
lower the free energy.

Let us consider a case when a two-band superconductor in a normal
state ($a_{1}a_{2}>\varepsilon^{2}$) has contact with a metal in a
superconducting state. Let the superconductor in a normal state
occupies a halfspace $x>0$. Since in a normal region the order
parameters are small, then minimization of the free energy
functional (\ref{1.1}) with respect to the order parameters gives
\begin{equation}\label{1.4}
\left\{\begin{array}{c}
  \frac{\hbar^{2}}{4m_{1}}\frac{d^{2}\Psi_{1}}{dx^{2}}+\frac{\hbar^{2}}{4}\eta\frac{d^{2}\Psi_{2}}{dx^{2}}-a_{1}\Psi_{1}-\varepsilon\Psi_{2}=0 \\
  \\
  \frac{\hbar^{2}}{4m_{2}}\frac{d^{2}\Psi_{2}}{dx^{2}}+\frac{\hbar^{2}}{4}\eta\frac{d^{2}\Psi_{1}}{dx^{2}}-a_{2}\Psi_{2}-\varepsilon\Psi_{1}=0 \\
\end{array}\right\}
\end{equation}
Eq.(\ref{1.4}) are a set of linear equations with constant
coefficients. Hence we must seek a solution in a form
$\Psi_{1}=\psi_{1}e^{kx}$, $\Psi_{2}=\psi_{2}e^{kx}$, where the
quantity $k$ has physical sense of an inverse coherence length:
$k=1/\xi$. Then we have
\begin{equation}\label{1.4a}
\left\{\begin{array}{c}
  \left(\frac{\hbar^{2}k^{2}}{4m_{1}}-a_{1}\right)\psi_{1}+\left(\frac{\hbar^{2}k^{2}}{4}\eta-\varepsilon\right)\psi_{2}=0 \\
  \\
  \left(\frac{\hbar^{2}k^{2}}{4}\eta-\varepsilon\right)\psi_{1}+\left(\frac{\hbar^{2}k^{2}}{4m_{2}}-a_{2}\right)\psi_{2}=0 \\
\end{array}\right\}.
\end{equation}
The characteristic equation are
\begin{equation}\label{1.5}
    k^{4}\left(\frac{\hbar^{2}}{4}\right)^{2}\left(\frac{1}{m_{1}m_{2}}-\eta^{2}\right)
    -k^{2}\frac{\hbar^{2}}{4}\left(\frac{a_{2}}{m_{1}}+\frac{a_{1}}{m_{2}}-2\eta\varepsilon\right)
    +a_{1}a_{2}-\varepsilon^{2}=0.
\end{equation}
Solutions of this equation corresponds to two coherence lengths.
At $T\rightarrow T_{c}$ we have
\begin{eqnarray}
  k_{1}^{2} &=& \frac{a_{1}a_{2}-\varepsilon^{2}}
  {\frac{\hbar^{2}}{4}\left(\frac{a_{2}}{m_{1}}+\frac{a_{1}}{m_{2}}-2\eta\varepsilon\right)}\label{1.5a} \\
  k_{2}^{2} &=& \frac{\left(\frac{a_{2}}{m_{1}}+\frac{a_{1}}{m_{2}}-2\eta\varepsilon\right)}
  {\frac{\hbar^{2}}{4}\left(\frac{1}{m_{1}m_{2}}-\eta^{2}\right)},\quad \eta^{2}\neq\frac{1}{m_{1}m_{2}}.\label{1.5b}
\end{eqnarray}
The first of them is $k_{1}=0$ at the critical temperature (when
$a_{1}a_{2}-\varepsilon^{2}=0$). That is the coherence length
$\xi_{1}=1/|k_{1}|$ is diverging $\xi_{1}(T\rightarrow
T_{c})\rightarrow\infty$. On the contrary $k_{2}(T=T_{c})\neq 0$
and it varies little with temperature. These length scales are not
related to the concrete bands involved in the formation of the
superconducting ordering in a system with interband interaction.
This result corresponds to the results in works \cite{litak,ord}
obtained by microscopic approach, however they suggested that the
intergradient interaction is absent ($\eta=0$).

According to the method for solving of a set of linear
differential equations with constant coefficients we have to write
solutions of Eq.(\ref{1.4}) in a form
\begin{equation}\label{1.5c}
    \begin{array}{c}
      \Psi_{1}=C_{1}\psi_{1}^{(1)}e^{k_{1}x}+C_{2}\psi_{1}^{(2)}e^{k_{2}x}\\
      \Psi_{2}=C_{1}\psi_{2}^{(1)}e^{k_{1}x}+C_{2}\psi_{2}^{(2)}e^{k_{2}x} \\
    \end{array},
\end{equation}
where coefficients $\psi_{1}^{(1)},\psi_{2}^{(1)}$ correspond to
the eigenvalue $k_{1}$ (they must be found from Eq.(\ref{1.4a})
substituting $k=k_{1}$), the coefficients
$\psi_{1}^{(2)},\psi_{2}^{(2)}$ correspond to the eigenvalue
$k_{2}$. Solutions (\ref{1.5c}) corresponds to boundary conditions
$\Psi_{1,2}(x\rightarrow\infty)=0$, that is $k_{1},k_{2}<0$. A
case $\eta^{2}>1/m_{1}m_{2}$, when the eigenvalue $k_{2}$ is
complex (the solution $\Psi=e^{k_{2}x}$ is oscillating), will be
considered below. From the first equation of Eq.(\ref{1.4a}) we
have:
\begin{equation}\label{1.5d}
    \psi_{2}=-\frac{\frac{\hbar^{2}k^{2}}{4m_{1}}-a_{1}}{\frac{\hbar^{2}k^{2}}{4}\eta-\varepsilon}\psi_{1}
\end{equation}
For $k=k_{1}=0$ (at $T=T_{c}$) we have
\begin{equation}\label{1.5e}
    \psi_{2}^{(1)}=-\frac{a_{1}}{\varepsilon}\psi_{1}^{(1)}.
\end{equation}
Eq.(\ref{1.5e}) conserves the phase relations (\ref{1.3a}): if
$\varepsilon<0$ the condensates in different bands are in a phase,
if $\varepsilon>0$ the condensates in different bands are in
antiphase (we are in a temperature region $T_{c1},T_{c2}<T<T_{c}$
hence $a_{1},a_{2}>0$). For $k=k_{2}$ and taking into account the
condition Eq.(\ref{1.3i}), which lowers the free energy of a
superconductor in a magnetic field, we have:
\begin{equation}\label{1.5i}
    \psi_{2}^{(2)}=-\frac{1}{\eta m_{1}}\frac{a_{2}m_{2}+2|\eta\varepsilon|m_{1}m_{2}+m_{1}^{2}m_{2}\eta^{2}a_{1}}
    {a_{2}m_{2}+a_{1}m_{1}+|\eta\varepsilon|m_{1}m_{2}+|\varepsilon/\eta|}\psi_{1}^{(2)}.
\end{equation}
In the case when the drag-effect is neglected $\eta=0$ we have:
\begin{equation}\label{1.5j}
    \psi_{2}^{(2)}=\frac{a_{2}m_{2}}{\varepsilon
    m_{1}}\psi_{1}^{(2)}.
\end{equation}
We can see that Eqs.(\ref{1.5i},\ref{1.5j}) are opposite to the
phase relations (\ref{1.3a}): when $\varepsilon<0$ then
$\psi_{2}^{(2)}=const\cdot\psi_{1}^{(2)}, const<0$ (because
$\eta>0$), for $\varepsilon>0$ it is analogously. This fact leads
to instability of a superconducting state in a spatial
inhomogeneous medium: any spatial inhomogeneity violates the phase
relations (\ref{1.3a}) and, consequently, suppress the
superconducting state. For some quantities of $\eta$ in a case
$\eta\varepsilon>0$ the dynamical stability can perhaps exist,
however in this case the London penetration depth (\ref{1.3e})
increases and, hence, the free energy (\ref{1.3h}) increases
compared with a case $\eta\varepsilon<0$. In a case
$\eta^{2}>1/m_{1}m_{2}$ the solution $\Psi=e^{k_{2}x}$ is
oscillating and it does not satisfy the boundary conditions
$\Psi_{1,2}(x\rightarrow\infty)=0$. The solution $\Psi=e^{k_{2}x}$
could be removed supposing $C_{2}=0$. However the eigenvalues
$k_{1}$ and $k_{2}$ are derived from the intrasystem interaction
and corresponds to the different length scales in the system.
Consequently their selection by the boundary conditions is
unphysical (unlike symmetric solutions $k$ and $-k$ one of which
can be selected according to the boundary conditions). Thus, to
ensure stability of a superconducting state and minimality of the
free energy, the solution $k_{1}$ must exist only. Then from
Eq.(\ref{1.5}) and Eqs.(\ref{1.3e},\ref{1.3h}) we can see that the
coefficient of intergradient interaction must be
\begin{equation}\label{1.6}
    \eta^{2}=\frac{1}{m_{1}m_{2}}, \quad \eta\varepsilon<0.
\end{equation}
In this case we have only one eigenvalue $k=k_{1}$ such that
$k(T\rightarrow T_{C})=0$.

Usind Eqs.(\ref{1.5a},\ref{1.6}) and Eq.(\ref{1.3}) we can obtain
the coherence length as
\begin{equation}\label{1.7}
    \xi^{2}=\frac{\frac{\hbar^{2}}{4}\left(\frac{a_{2}}{m_{1}}+\frac{a_{1}}{m_{2}}+2|\eta||\varepsilon|\right)}{|a_{1}a_{2}-\varepsilon^{2}|}
    \approx\sqrt{\frac{a_{2}}{a_{1}}}\frac{\frac{\hbar^{2}}{4}\left(\frac{1}{m_{1}}+\frac{a_{1}}{a_{2}m_{2}}+2|\eta|\sqrt{\frac{a_{1}}{a_{2}}}\right)}
    {2\left|\sqrt{a_{1}a_{2}}-|\varepsilon|\right|},
\end{equation}
At the critical temperature this coherence length diverges
$\xi(T\rightarrow T_{c})\rightarrow\infty$ because
$a_{1}(T_{c})a_{2}(T_{c})-\varepsilon^{2}=0$. The similar problem,
junction between a single-band superconductor and a two-band
superconductor, was considered in \cite{tan}, where it is assumed
the phase shift $\Delta\theta$ in the junction is zero under the
condition of no current and no field. However the current through
the junction between two superconductors is
$J=J_{0}\sin\Delta\theta$, that is the condition $J=0$ is
satisfied by both $\Delta\theta=0$ and $\Delta\theta=\pi$. In
\cite{brink} it has been shown the dependence of the current on
the phase difference $J\propto\sin\Delta\theta$ for the junction
between a single-band superconductor and a two-band superconductor
also takes place. Since the phase different in a two-band
superconductor is either $0$ or $\pi$ - Eq.(\ref{1.3a}), then
proximity of a single-band superconductor can not change the phase
relation in a two-band superconductor.

Single coherence length allows us to represent the orders
parameters in a form
$\Psi_{2}(\textbf{r})=C(T)\Psi_{1}(\textbf{r})$, where the
coefficient $C$ is not function of spatial coordinates (as follows
from the above, $C>0$ if $\varepsilon<0$ and $C<0$ if
$\varepsilon>0$). Hence the free energy functional of a two-band
superconductor (\ref{1.1}) can be rewritten in the form of GL
functional of a single-band superconductor:
\begin{eqnarray}\label{1.8}
   F=\int d^{3}r\left[\frac{\hbar^{2}}{4M}\left|\left(\nabla-\frac{2\pi i}{\Phi_{0}}\textbf{A}\right)\Psi\right|^{2}
   +A\left|\Psi\right|^{2}+\frac{B}{2}\left|\Psi\right|^{4}+\frac{H^{2}}{8\pi}\right],
\end{eqnarray}
where the coefficients have a form
\begin{eqnarray}
   &&A=a_{1}+a_{2}C^{2}+2\varepsilon C\label{1.9}\\
   &&B=b_{1}+b_{2}C^{4}\label{1.10}\\
   &&M^{-1}=\frac{1}{m_{1}}+\frac{C^{2}}{m_{2}}+\frac{2|C|}{\sqrt{m_{1}m_{2}}},\label{1.11}
\end{eqnarray}
and we have redesignated $\Psi\equiv\Psi_{1}$. Thus the theory of
a two-band superconductor is reduced to GL theory of a single-band
superconductor. All characteristics (coherence length, magnetic
penetrations depth, GL parameter, critical magnetic fields,
magnetization, critical currents in a wire etc.) can be found by
usual GL theory. However, unlike GL theory, the coefficient $B$
and the effective mass $M$ are functions of temperature since the
coefficient $C$ is a function of temperature.

Now we should find the coefficient $C$. Let us substitute
$\Psi_{2}=C\Psi_{1}$ in Eq.(\ref{1.2}):
\begin{equation}\label{1.12}
\left\{\begin{array}{c}
  a_{1}+\varepsilon C+b_{1}\Psi_{1}^{2}=0 \\
  a_{2}C+\varepsilon+b_{2}C^{3}\Psi_{1}^{2}=0 \\
\end{array}\right\}
\end{equation}
If $T\rightarrow T_{c}$ then the equations can be linearized. In
this case we have solutions $C=-a_{1}/\varepsilon$ or
$C=-\varepsilon/a_{2}$. Near the critical temperature we can use
Eq.(\ref{1.3}), that is $|\varepsilon|=\sqrt{a_{1}a_{2}}$. Then
the solution becomes unique:
\begin{equation}\label{1.13}
    \begin{array}{cc}
      C=\sqrt{\frac{a_{1}}{a_{2}}}, & \texttt{if} \quad\varepsilon<0 \\
      C=-\sqrt{\frac{a_{1}}{a_{2}}}, & \texttt{if} \quad\varepsilon>0 \\
    \end{array}
\end{equation}
This approximation expresses the fact that relation between the
order parameters is determined by the single-band critical
temperatures $T_{c1}, T_{c2}$: if $T_{c1}>T_{c2}$ then
$\Delta_{1}>\Delta_{2}$ - Fig.(\ref{Fig1}).

Using Eqs.(\ref{1.9},\ref{1.10},\ref{1.11},\ref{1.12}) we can find
main characteristics of a superconductor as in the usual GL
theory. A coherence length:
\begin{equation}\label{1.14}
    \xi^{2}=\frac{\hbar^{2}}{4M|A|}=\frac{\hbar^{2}}{4}\frac{
    \left(\frac{1}{m_{1}}\sqrt{\frac{a_{2}}{a_{1}}}+\frac{1}{m_{2}}\sqrt{\frac{a_{1}}{a_{2}}}+\frac{2}{\sqrt{m_{1}m_{2}}}\right)}
    {2\left|\sqrt{a_{1}a_{2}}-|\varepsilon|\right|},
\end{equation}
a magnetic penetrations depth:
\begin{equation}\label{1.15}
    \lambda^{2}=\frac{Mc^{2}B}{8\pi e^{2}|A|}=\frac{c^{2}}
    {8\pi e^{2}}
    \frac{b_{1}\frac{a_{2}}{a_{1}}+b_{2}\frac{a_{1}}{a_{2}}}
    {2\left|\sqrt{a_{1}a_{2}}-|\varepsilon|\right|
    \left(\frac{1}{m_{1}}\sqrt{\frac{a_{2}}{a_{1}}}+\frac{1}{m_{2}}\sqrt{\frac{a_{1}}{a_{2}}}+\frac{2}{\sqrt{m_{1}m_{2}}}\right)},
\end{equation}
a GL parameter:
\begin{equation}\label{1.16}
    \kappa=\frac{\lambda}{\xi}=\frac{c}{2\sqrt{\pi} e}M\sqrt{B}=\frac{c}{2\sqrt{\pi} e}
    \frac{\sqrt{b_{1}\frac{a_{2}}{a_{1}}+b_{2}\frac{a_{1}}{a_{2}}}}
    {\left(\frac{1}{m_{1}}\sqrt{\frac{a_{2}}{a_{1}}}+\frac{1}{m_{2}}\sqrt{\frac{a_{1}}{a_{2}}}+\frac{2}{\sqrt{m_{1}m_{2}}}\right)},
\end{equation}
We can see that the GL parameter is a function of temperature
unlike single-band GL theory. However this dependence is type
$\frac{T-T_{c1}}{T-T_{c2}}$ that is little varying function of
temperature if $T\gg T_{c1},T_{c2}$. It should be noticed that
this approximation is correct if $T> T_{c1},T_{\texttt{c}2}$ only.
We can extrapolate the obtained expressions for all temperature.
To do it we can suppose $a_{i}=\gamma_{i}(T_{c}-T_{ci})=const$,
then $M=const,B=const$, however it is necessary to expand the
expression $\sqrt{a_{1}a_{2}}-|\varepsilon|$ in powers of
$T-T_{c}$:
\begin{eqnarray}\label{1.17}
    &&\sqrt{a_{1}a_{2}}-|\varepsilon|=\frac{\gamma_{1}\gamma_{2}\left(2T_{c}-T_{c1}-T_{c2}\right)}{2\sqrt{a_{1}a_{2}}}\left(T-T_{c}\right)\nonumber\\
    &&+\frac{\gamma_{1}\gamma_{2}}{2\sqrt{a_{1}a_{2}}}
    \left(1-\frac{\gamma_{1}\gamma_{2}\left(2T_{c}-T_{c1}-T_{c2}\right)^{2}}{4a_{1}a_{2}}\right)\left(T-T_{c}\right)^{2}+\ldots
\end{eqnarray}
Thus in the functional for a two-band superconductor the
coefficient $A$ is a power series of $\left(T-T_{c}\right)$ unlike
the GL functional for a single-band superconductor. From this fact
a nonlinear temperature dependence of the upper critical field
follows (hear $f_{1},f_{2}$ are some coefficients):
\begin{equation}\label{1.18}
    H_{c2}=\frac{\Phi_{0}}{2\pi\xi^{2}}\propto\begin{array}{cc}
       \left(T_{c}-T\right) & \quad\texttt{single-band GL theory} \\
       f_{1}\left(T_{c}-T\right)+f_{2}\left(T_{c}-T\right)^{2}+\ldots & \quad\texttt{two-band GL theory} \\
    \end{array},
\end{equation}
that is consistent with experimental data (in bulk
$\texttt{LuNi}_{2}\texttt{B}_{2}\texttt{C}$, $\texttt{MgB}_{2}$)
in \cite{freu,mull,bud} and theoretical results in
\cite{asker0,asker1}, where it has been shown that the presence of
two order parameters for two bands yields a nonlinear temperature
dependence of $H_{c2}(T)$ in the vicinity of the critical
temperature unlike the single-band s-wave BCS theory and GL
theory. It should be noted that this difference can be a cause of
strong enhancement of $H_{c2}(T)$ (up to ten-fold increase) in
dirty two-gap superconductors, that, as noted in \cite{gur2}, is
result from the anomalous upward curvature of $H_{c2}(T)$. For the
lower critical field $H_{c1}$ and the thermodynamic magnetic field
$H_{cm}$ we have analogous expansion because
\begin{eqnarray}
    &&H_{c1}=\frac{\Phi_{0}}{2\pi\lambda^{2}}\ln\kappa\propto|\varepsilon|-\sqrt{a_{1}a_{2}}\label{1.19}\\
    &&H_{cm}=\frac{\Phi_{0}}{2\sqrt{2}\pi\lambda\xi}\propto|\varepsilon|-\sqrt{a_{1}a_{2}}\label{1.20},
\end{eqnarray}
that demonstrates nonlinear temperature dependence too and
correlates with theoretical results of \cite{asker4,asker1}. Let
the carriers have different effective masses in different bands,
for example $m_{1}\gg m_{2}$. From Eq.(\ref{1.11}) we can see that
the two-band effective mass $M$ is determined mainly by the
smaller mass $m_{2}$. From Eqs.
(\ref{1.14},\ref{1.15},\ref{1.18},\ref{1.19}) we can see the
critical fields $H_{c1}$ and $H_{c2}$ depend on the effective mass
as $H_{c1}\propto 1/M$, $H_{c2}\propto M$. Hence, as noted in
\cite{asker1,asker2,asker3}, the critical fields are determined
mainly by the smaller mass $m_{2}$, while the contribution from
the lager mass can be neglected.

\section{Multi-band superconductor}\label{multi}

Using results of previous section we can generalize the
above-described method for two-band superconductors to multi-band
superconductors. In presence of n order parameters in an isotropic
s-wave superconductor, the free energy functional can be written
as
\begin{eqnarray}\label{2.1}
    &&F=\int d^{3}r[\sum_{i=1}^{n}[\frac{\hbar^{2}}{4m_{i}}\left|D\Psi_{i}\right|^{2}
    +a_{i}\left|\Psi_{i}\right|^{2}+\frac{b_{i}}{2}\left|\Psi_{i}\right|^{4}\nonumber\\
    &&+\sum_{j=2,j>i}^{n}
    \frac{\hbar^{2}}{4}\eta_{ij}\left(D^{+}\Psi_{i}^{+}D\Psi_{j}
    +D\Psi_{i}D^{+}\Psi_{j}^{+}\right)\nonumber\\
    &&+\sum_{j=2,j>i}^{n}\varepsilon_{ij}\left(\Psi_{i}^{+}\Psi_{j}+\Psi_{i}\Psi_{j}^{+}\right)]+\frac{H^{2}}{8\pi}],
\end{eqnarray}
Critical temperature can be found from an equation:
\begin{equation}\label{2.2}
    \left|%
\begin{array}{cccc}
  a_{1} & \varepsilon_{12} & \ldots & \varepsilon_{1n} \\
  \varepsilon_{12} & a_{2} & \ldots & \varepsilon_{2n}\\
  \ldots & \ldots & \ldots & \ldots \\
  \varepsilon_{1n} & \varepsilon_{2n} & \ldots & a_{n} \\
\end{array}
\right|=0,
\end{equation}
which is analog of Eq.(\ref{1.3}). However we should notice that
in general case the symmetry $\varepsilon\leftrightarrow
-\varepsilon$ for critical temperature, like in the two-band case,
is absent. If all $\varepsilon_{ij}<0$ but some
$\varepsilon_{ij}>0$, suppression of superconductivity is
possible. We will consider a case of attractive interband
interaction only, that is all $\varepsilon_{ij}<0$.

Following our scheme we should find coefficients of the
intergradients interaction $\eta_{ij}$ and the coherence length
$\xi$. Equation for the coherence length $\xi^{2}=1/k^{2}$ is
\begin{eqnarray}\label{2.3}
    &&\left|%
\begin{array}{cccc}
  \frac{\hbar^{2}}{4m_{1}}k^{2}-a_{1} & \frac{\hbar^{2}}{4}\eta_{12}k^{2}-\varepsilon_{12} & \ldots & \frac{\hbar^{2}}{4}\eta_{1n}k^{2}-\varepsilon_{1n} \\
  \frac{\hbar^{2}}{4}\eta_{12}k^{2}-\varepsilon_{12} & \frac{\hbar^{2}}{4m_{2}}k^{2}-a_{2} & \ldots & \frac{\hbar^{2}}{4}\eta_{2n}k^{2}-\varepsilon_{2n}\\
  \ldots & \ldots & \ldots & \ldots \\
  \frac{\hbar^{2}}{4}\eta_{1n}k^{2}-\varepsilon_{1n} & \frac{\hbar^{2}}{4}\eta_{2n}k^{2}-\varepsilon_{2n} & \ldots & \frac{\hbar^{2}}{4m_{n}}k^{2}-a_{n} \\
\end{array}
\right|\nonumber\\
&&\nonumber\\
&&=f_{n}(m_{i},\eta_{ij})k^{2n}+f_{n-1}(m_{i},\eta_{ij})k^{2(n-1)}+\ldots+f_{1}(m_{i},\eta_{ij})k^{2}+f_{0}=0,
\end{eqnarray}
which is analog of Eq.(\ref{1.5}). At $T=T_{c}$ we have $f_{0}=0$.
The coefficients $\eta_{ij}$ must be such that the functions
$f_{n}=f_{(n-1)}=\ldots=f_{2}=0$, then the coherence length is
\begin{equation}\label{2.4}
    1/\xi^{2}=k^{2}=-\frac{f_{0}}{f_{1}}.
\end{equation}
By analogy of (\ref{1.6}) and using the condition
$\varepsilon_{ij}<0$ we can suppose
\begin{equation}\label{2.4a}
    \eta_{ij}=\frac{1}{\sqrt{m_{i}m_{j}}}.
\end{equation}

In the next step we should to represent the orders parameters in a
form
$\Psi_{2}=C_{2}(T)\Psi_{1},\Psi_{3}=C_{3}(T)\Psi_{1},\ldots,\Psi_{n}=C_{n}(T)\Psi_{1}$.
Then the free energy functional of a multi-band superconductor
(\ref{2.1}) takes the form of the GL functional (\ref{1.8}) of a
single-band superconductor with coefficients
\begin{eqnarray}
   &&A=a_{1}+\sum_{i=2}^{n}a_{i}C_{i}^{2}+2\sum_{i=2}^{n}\varepsilon_{1i}C_{i}+2\sum_{i=2}^{n}\sum_{j=3,j>i}^{n}\varepsilon_{ij}C_{i}C_{j}\label{2.5}\\
   &&B=b_{1}+\sum_{i=2}^{n}b_{i}C_{i}^{4}\label{2.6}\\
   &&M^{-1}=\frac{1}{m_{1}}+\sum_{i=2}^{n}\frac{C_{i}^{2}}{m_{i}}+2\sum_{i=2}^{n}\eta_{1i}C_{i}+2\sum_{i=2}^{n}\sum_{j=3,j>i}^{n}\eta_{ij}C_{i}C_{j}\label{2.7}
\end{eqnarray}
Linearized equations for $C_{2},C_{3},\ldots,C_{n}$ are
\begin{equation}\label{2.8}
\left\{\begin{array}{c}
  a_{1}+\varepsilon_{12} C_{2}+...\varepsilon_{1n} C_{n}=0 \\
  \varepsilon_{12}+a_{2}C_{2}+...\varepsilon_{2n} C_{n}=0 \\
  \ldots \\
  \varepsilon_{1n}+\varepsilon_{23}C_{2}+...a_{n}C_{n}=0\\
\end{array}\right\}
\end{equation}
which have to be solved taking into account Eq.(\ref{2.2}) so that
the solutions are unequivocal (as we have shown in the two-band
case). However we can use an approximate method. In the two-band
problem we supposed the coefficient
$C=\sqrt{\frac{a_{1}}{a_{2}}}=\sqrt{\frac{\gamma_{1}}{\gamma_{2}}\frac{T-T_{c1}}{T-T_{c2}}}$
for $\Psi_{2}=C\Psi_{1}$, that is relation between the order
parameters is determined by the single-band critical temperatures
$T_{c1}, T_{c2}$: if $T_{c1}>T_{c2}$ then $\Delta_{1}>\Delta_{2}$.
This fact can be used for the coefficients $C_{i}$ in the
multi-band problem, where we can suppose:
\begin{equation}\label{2.9}
    C_{2}=\sqrt{\frac{a_{1}}{a_{2}}},\quad
    C_{3}=\sqrt{\frac{a_{1}}{a_{3}}},\ldots,\quad
    C_{n}=\sqrt{\frac{a_{1}}{a_{n}}}.
\end{equation}
Substituting Eq.(\ref{2.9}) in Eq.(\ref{2.5}) and reducing to a
common denominator we have
\begin{equation}\label{2.10}
    A=\frac{n\sqrt{a_{1}}}{\prod_{i=2}^{n}\sqrt{a_{i}}}f(a_{i},\varepsilon_{ij}),
\end{equation}
where
\begin{equation}\label{2.11}
    f(a_{i},\varepsilon_{ij})=\prod_{i=1}^{n}\sqrt{a_{i}}+\frac{2}{n}\sum_{i=2}^{n}\varepsilon_{1i}\prod_{k=2,k\neq i}^{n}\sqrt{a_{k}}
    +\frac{2\sqrt{a_{1}}}{n}\sum_{i=2}^{n}\sum_{j=3,j>i}^{n}\varepsilon_{ij}\prod_{k=2,k\neq i,k\neq j}^{n}\sqrt{a_{k}},
\end{equation}
The critical temperature is such a temperature when
$f(T=T_{c})=0$. As in the two-band problem we can extrapolate the
obtained expressions for all temperature. To do this we have to
suppose $a_{i}=\gamma_{i}(T_{c}-T_{ci})=const$, then
$M=const,B=const$, however it is necessary to expand the
expression $f(a_{i},\varepsilon_{ij})$ in powers of $T-T_{c}$.
Thus the multi-band  problem is reduced to the single-band problem
with the effective mass $M$, however the coefficient $A$ is power
series of $(T-T_{c})$ unlike the GL free energy functional.

\section{Results}\label{concl}

In this work we have shown that the term of the drag effect
$\eta\left(\nabla\Psi_{1}^{+}\nabla\Psi_{2}+\nabla\Psi_{1}\nabla\Psi_{2}^{+}\right)$
in the free energy functional of an isotropic bulk two-band
superconductor plays important role and the restrictions for the
coefficient $\eta$ exist. If the coefficient is
$\eta^{2}=\frac{1}{m_{1}m_{2}}$ and it's sign is opposite to the
sign of the coefficient in the term of the proximity effect
$\varepsilon\left(\Psi_{1}^{+}\Psi_{2}+\Psi_{1}\Psi_{2}^{+}\right)$,
that is $\eta\varepsilon<0$, then this leads to a single coherence
length $\xi$, which diverges at the critical temperature
$\xi(T\rightarrow T_{c})\rightarrow\infty$, and a single GL
parameter. This quantity ensures the stability of a superconductor
state and the least possible free energy in this case. Other
quantities of the coefficient or neglecting of the drag effect
$\eta=0$ leads, at first, to the existence of two coherence
lengths, where one of them diverges at the critical temperature
while the second length is finite at all temperatures. Secondly,
it leads to the dynamical instability (suppressing of a
superconducting state if the order parameters are spatial
inhomogeneous) due to violation of the phase relations
(\ref{0.2},\ref{1.3a}). These results mean that the isotropic bulk
type-1.5 superconductors are impossible.

It should be noticed that these results are obtained in the GL
domain only. Hence it can be supposed that at low temperatures the
disproportion $\Psi_{2}(\textbf{r},T)\neq
C(T)\Psi_{1}(\textbf{r},T)$ can takes place, that is there are two
different coherence lengths $\xi_{1}\neq\xi_{2}$. However this
fact means that the order parameters have different gradients
$\nabla\Psi_{1}(\textbf{r})\neq \nabla\Psi_{2}(\textbf{r})$. Since
the order parameters are $|\Psi_{1}|e^{i\varphi_{1}},
|\Psi_{2}|e^{i\varphi_{2}}$, then the different gradients can lead
to violation of the equality (\ref{1.3c}), hence to violation of
the phase relations (\ref{1.3a}). Thus the state with different
coherence lengths is dynamically unstable.

The approximate method for solving of set of GL equations for an
isotropic bulk multi-band superconductor has been developed. Using
the results about the drag effect we have shown that the free
energy functional for a two-band superconductor can be reduced to
the GL functional for an effective single-band superconductor.
This effective superconductor is characterized with some an
effective mass of carriers (as a function of $m_{1},m_{2},\eta$)
and a coefficient at $|\Psi|^{2}$ as a power series of $(T-T_{c})$
in the vicinity of the critical temperature. This temperature
dependence causes nonlinear dependence of upper and lower critical
fields $H_{c2},H_{c1}$, thermodynamical magnetic fields $H_{cm}$
on temperature unlike the single-band GL theory. Generalizing this
result we have developed an algorithm which allows to reduce the
free energy functional of a multi-band superconductor to the
effective GL free energy functional of a single-band
superconductor provided that all interband interactions are
attractive.

\acknowledgments

The work is supported by the project \#0112U000056 of the National
Academy of Sciences of Ukraine.

\end{document}